\documentclass{llncs}
\usepackage{makeidx}  
\usepackage{amsfonts}
\usepackage{amsmath}
\usepackage[dutch,USenglish]{babel}
\begin{document}

\title{On the Complexity of Several Haplotyping Problems}
\titlerunning{On the complexity of several haplotyping problems}  
%
\author{Rudi Cilibrasi\inst{2}\thanks{Research of all authors paid by the Dutch
BSIK-Bricks project AFM2.}, Leo van Iersel\inst{1}, Steven Kelk\inst{2} \and John Tromp\inst{2}}

%
\authorrunning{Rudi Cilibrasi, Leo van Iersel, Steven Kelk and John Tromp}   
%
%
\institute{Technische Universiteit Eindhoven (TU/e), Den Dolech 2, 5612 AX Eindhoven, Netherlands,\\
\email{l.j.j.v.iersel@tue.nl},\\
\texttt{http://w3.tue.nl/nl/}
\and
Centrum voor Wiskunde en Informatica (CWI), Kruislaan 413, 1098 SJ Amsterdam, Netherlands, \\
\email{Rudi.Cilibrasi@cwi.nl, S.M.Kelk@cwi.nl, John.Tromp@cwi.nl}, \\
\texttt{http://www.cwi.nl}
}

\maketitle              

\begin{abstract}
In this paper we present a collection of results pertaining to
haplotyping. The first set of results concerns the combinatorial problem
of reconstructing haplotypes from incomplete and/or imperfectly
sequenced haplotype data. More specifically, we show that an interesting, restricted case of
\emph{Minimum Error Correction} (MEC)
is NP-hard, point out problems in earlier claims about a related
problem, and present a polynomial-time algorithm for
the ungapped case of \emph{Longest Haplotype Reconstruction} (LHR). Secondly, we
present a polynomial time algorithm for the problem of
resolving genotype data using as few haplotypes as possible (the
\emph{Pure Parsimony Haplotyping Problem}, PPH) where each genotype has at
most two ambiguous positions, thus solving an open problem posed by Lancia et al in \cite{pureparsimony}.
\end{abstract}
\section{Introduction}
If we abstractly consider the human genome as a string over the nucleotide alphabet $\{ A, G, C, T \}$, it is widely
known that the genomes of any two humans are more than 99\% similar. In other words, it is known that,
at most sites along the genome, humans all have the same nucleotide. At certain specific sites along the
genome, however, variability is observed across the human population. These sites are known as \emph{Single
Nucleotide Polymorphisms} (SNPs) and are formally defined as the sites on the human genome where, across
the human population, two or more nucleotides are observed and each such nucleotide occurs in at
least 5\% of the population. It turns out that these sites, which occur (on average) approximately once per
thousand bases of the human genome, capture the bulk of human genetic variability; the string of nucleotides found at the SNP sites
of a human - the \emph{haplotype} of that individual - can thus be thought of as a ``fingerprint'' for that individual.
It is further apparent that, for most SNP sites, only two nucleotides are seen; sites where three or more nucleotides are possible
are comparatively rare. Thus, from a combinatorial perspective, a haplotype can be abstractly expressed as a string
over the alphabet $\{ 0,1 \}$. Indeed, the biologically-motivated field of SNP and haplotype analysis - which
is at the forefront of ``real-world'' bioinformatics - has spawned an impressively rich and varied assortment of combinatorial
problems, which are well described in surveys such as \cite{bonizzoni} and \cite{halldorsson}. In this paper we focus on three such
combinatorial problems; the first two are related to the problem of haplotyping a single individual,
and the third is related to the problem of explaining the genetic variability of a population using as few
haplotypes as possible.\\
\\
The first two problems are both variants of the \emph{Single Individual Haplotyping Problem} (SIH), introduced
in \cite{lanciabafna}. The SIH
problem amounts to determining the haplotype of an individual
using (potentially) incomplete and/or imperfect fragments of sequencing data. The situation is further complicated by the fact
that, being a \emph{diploid} organism, a human has two versions of each chromosome; one each
from the individual's mother and father. Hence, for a given interval of the genome, a human actually
has two haplotypes. Thus, the SIH problem can be more
accurately described as finding the two haplotypes of an individual given fragments of sequencing data
where the fragments potentially have read errors and, crucially, where it is \emph{not} known which of the
two chromosomes each fragment was read from. There are four well-known variants
of the problem: \emph{Minimum Fragment Removal} (MFR), \emph{Minimum SNP Removal} (MSR), \emph{Minimum Error Correction} (MEC),
and \emph{Longest Haplotype Reconstruction} (LHR). In this paper we give results for MEC and LHR and refer
the reader to \cite{bafna2005} for information about MFR and MSR.
\subsection{Minimum Error Correction (MEC)}
\label{intro:mec}
This is the problem where the input is an $n \times m$ matrix $M$ of SNP fragments.
Each column of $M$ represents an SNP site and thus each element of the matrix denotes the (binary) choice
of nucleotide seen at that SNP location
on that fragment. An element of the matrix can thus either be `0', `1' or a \emph{hole},
represented by `-', which denotes lack of knowledge or uncertainty about the nucleotide at that site.
We use $M[i,j]$ to refer to the value found at row $i$, column $j$ of $M$, and
use $M[i]$ to refer to the $i$th row. We say that two rows $r_1, r_2$ of the matrix are in \emph{conflict} if there
exists a column $j$ such that $M[r_1, j] \neq M[r_2, j]$ and $M[r_1,j], M[r_2, j] \in \{0,1\}$. We say that a matrix is
\emph{feasible} if the rows of the matrix can be partitioned into two sets such that all rows within each set are pairwise
non-conflicting. The goal with MEC is thus to ``correct'' (or ``flip'') as few entries of the input matrix as possible
(i.e. convert 0 to 1 or vice-versa) to make the resulting matrix feasible. The motivation behind this is that
all rows of the input matrix were sequenced from one haplotype or the other, and that any deviation from
that haplotype occurred because of read-errors during sequencing.\\
\\
In the context of haplotyping, MEC has been discussed - sometimes under a different name - in
papers such as \cite{bonizzoni}, \cite{fasthare}, \cite{greenberg} and (implicitly) \cite{lanciabafna}.
One question arising from this discussion is how the distribution of holes in the input data affects
computational complexity. To explain,
let us first define a \emph{gap} (in a string over the alphabet $\{0,1,-\}$) as a maximal contiguous block of holes
that is flanked on both sides by non-hole values. For example, the string \texttt{---0010---} has no gaps, \texttt{-0--10-111} has
two gaps, and \texttt{-0-----1--} has one gap\footnote{The case where each row of the input matrix has at most 1 gap
is considered biologically relevant because \emph{double-barrelled shotgun sequencing} produces two disjoint intervals
of sequencing data.}. (Note that the presence of holes does not automatically imply the
presence of gaps!) The problem variant \emph{Ungapped-MEC} is where every row of the input matrix is ungapped i.e.
all holes appear at the start or end.\\
\\
In this paper we offer what we believe is the first concrete proof that Ungapped-MEC (and hence
the more general Gapped-MEC) is NP-hard. We do so by
reduction from the optimisation version of MAX-CUT. As far as
we are aware, other claims of this result are based explicitly or implicitly on results found in \cite{kleinberg}; as we discuss
in Section \ref{sec:mec}, we fear that the results in \cite{kleinberg} cannot be used for this purpose. Directly related to
this, we define the problem \emph{Binary-MEC}, where the input matrix contains no holes; as far as we know the
complexity of this problem is still - intriguingly - open.
\subsection{Longest Haplotype Reconstruction (LHR)}
\label{intro:lhr}
In this variant of the SIH problem, the input is again an SNP matrix $M$ with elements drawn from $\{0,1,-\}$. Recall
that the rows of a feasible matrix $M$ can be partitioned into two sets such that all rows within each set are
pairwise non-conflicting. Having obtained such a partition, we can reconstruct a haplotype from each set by merging all the rows in that set together. (We define this
formally later in Section \ref{sec:lhr}.) With LHR the goal is to remove \emph{rows} such that the resulting matrix is
feasible and such that the sum of the lengths of the two resulting haplotypes is maximised. In this
paper we show that \emph{Ungapped-LHR} (where ungapped is defined as before) is polynomial-time solvable and
we give a dynamic programming algorithm for this which runs in time $O(n^{2}m+n^{3})$ for an $n \times m$ input
matrix. This improves upon the result of \cite{lanciabafna}; the result of \cite{lanciabafna} also showed a polynomial-time
algorithm for Ungapped-LHR but under the restricting assumption of non-nested input rows.
\subsection{Pure Parsimony Haplotyping Problem (PPH)}
\label{intro:pphp}
As mentioned earlier, there are actually two haplotypes for any given interval of an individual's genome.
With current sequencing techniques it is still considered impractical to read the two haplotypes separately; instead,
a single string is returned - the \emph{genotype} - which combines the data from the two haplotypes but, in
doing so, loses some information. Thus, whereas a haplotype is a string over the $\{0,1\}$ alphabet,
a genotype is a string over the $\{0,1,2\}$ alphabet. A `0' (respectively, `1') entry in the genotype means that both chromosomes
have a `0' (respectively, `1') at that position. In contrast, a `2' entry means that the two haplotypes \emph{differ} at
that location: one has a `0' while the other has a `1' but we don't know which goes where. Thus, a `2'-site of a genotype is called
an \emph{ambiguous} position. We say that two haplotypes \emph{resolve} a given genotype if that genotype is
the result of combining the two haplotypes in the above manner. For example, the pair of haplotypes \texttt{0110} and \texttt{0011} resolve
the genotype \texttt{0212}.\\
\\
It follows that a genotype with $a \geq 1$ ambiguous positions can be resolved in $2^{a-1}$ ways. Now, suppose we have a population of individuals
and we obtain (without errors) the genotype of each individual. The \emph{Pure Parsimony Haplotyping Problem} (PPH) is
as follows:- given a set of genotypes, what is the smallest number of haplotypes such that each genotype
is resolved by some pair of the haplotypes? In \cite{pureparsimony} it is shown that PPH is hard (i.e. NP-hard and APX-hard)
even in the restricted case where no genotype has more than 3 ambiguous positions. The case of 2 ambiguous positions per
genotype is left as an open question in \cite{pureparsimony}. In this paper
we resolve this question by providing a polynomial-time algorithm for this problem that has
a running time of $O(mn\log(n) + n^{3/2})$ for $n$ genotypes each of length $m$.

\section{Minimum Error Correction (MEC)}
\label{sec:mec}

For a length-$m$ string $X \in \{0,1,-\}^m$, and a length-$m$ string $Y 
\in \{0,1\}^m$, we define
$d(X,Y)$ as being equal to the number of \emph{mismatches} between the strings i.e. positions where $X$ is 0 and $Y$ is 1,
or vice-versa. (Holes do not contribute to the mismatch count.) An $n \times m$ SNP matrix $M$ is \emph{feasible} if
there exist two strings (haplotypes) $H_1, H_2 \in \{0,1\}^m$, such that for all
rows $r \in M$, $d( r, H_1) = 0$ or $d( r, H_2 )=0$. A \emph{flip} is where a 0 entry is converted to a 1, or
vice-versa. Note that, in our formulation of the problem, we do not allow flipping to or from holes, and the haplotypes $H_1$ and $H_2$
may not contain holes.\\
\\
\textbf{Problem:} \emph{Ungapped-MEC}\\
\textbf{Input:} An ungapped SNP matrix $M$\\
\textbf{Output:} The smallest number of flips needed to make $M$ feasible.\\
\\
Note that Ungapped-MEC is an optimisation problem, not a decision problem, hence the use of ``NP-hard'' in
the following lemma rather than ``NP-complete''. A decision version may be obtained by adding a flip
upperbound in the range $[0,nm]$.

\begin{lemma}
Ungapped-MEC is NP-hard.
\end{lemma}
\begin{proof}
We give a polynomial-time Turing reduction from the optimisation version
of MAX-CUT, which is the problem of computing the size of a maximum
cut in a graph. Let $G=(V,E)$ be the input to MAX-CUT, where $E$ is undirected. (Without loss of generality we identify
$V$ with the natural numbers $1, 2, ..., |V|$.) We construct an instance $M$ of
Ungapped-MEC as follows. $M$ has $2k + |E|$ rows and $2|V|$ columns where $k = 2|E||V|^2$. We use $M_0$ to
refer to the first $k$ rows of $M$, $M_1$ to refer to the second $k$ rows of $M$, and $M_G$ to refer to the remaining $|E|$
rows. The first $k/|V|$ rows of $M_0$ all have the following pattern: a 0 in the first column, a 0
in the second column, and the rest of the row is holes. The second $k/|V|$ rows of $M_0$ all
have a 0 in the third column, a 0 in the fourth column, and the rest holes; we continue this
pattern i.e. each row in the $j$th block of $k/|V|$ rows in $M_0$ ($1 \leq j \leq |V|$) has a 0 in column $2j-1$, a 0 in
column $2j$, and the rest holes. $M_1$ is defined identically except that 1s are used instead
of 0s. Each row of $M_G$ encodes an edge from $E$:- for an edge $(i,j)$ (where $i$ is the numerically lower
endpoint) we specify that columns $2i-1$ and $2i$ contain 0s, columns $2j-1$ and $2j$ contain 1s, and for all $c \neq i, j$,
column $2c-1$ contains 0 and column $2c$ contains 1.\\
\\
Suppose $t$ is the largest cut possible in $G$. We claim that:\\
\begin{equation}
\label{maxcut}
Ungapped\text{-}MEC(M)= |E|(|V|-2) + 2(|E|-t)
\end{equation}
From this $t$ (i.e. MAX-CUT(G)) can easily be computed. First, note that the solution to Ungapped-MEC(M) is trivially upperbounded
by $|V||E|$. This follows because we could simply flip every 1 entry in $M_G$ to 0; the resulting overall matrix would be
feasible because we could just take $H_0$ as the all-0 string and $H_1$ as the
all-1 string. Now, we say a haplotype $H$ has the \emph{double-entry} property if, for all odd-indexed positions (i.e. columns) $j$ in $H$, the entry at position $j$ of $H$ is
the same as the entry at position $j+1$. We argue that a minimal number of feasibility-inducing flips will \emph{always} lead to
two haplotypes $H_1, H_2$ such that both haplotypes have the double-entry property and, further, $H_1$ is the bitwise complement
of $H_2$. (We describe such a pair of haplotypes as \emph{partition-encoding}.) This is because, if $H_1, H_2$ are not
partition-encoding, then at least $k/|V| > |V||E|$ (in contrast with zero) entries in $M_0$ and/or $M_1$ will have to be flipped, meaning this strategy is doomed to begin with.\\
\\	
Now, for a given partition-encoding pair of haplotypes, it follows that - for each row in $M_G$ - we will have to flip
either $|V|-2$ or $|V|$ entries to reach its nearest haplotype. This is because, irrespective of which
haplotype we move a row to, the $|V|-2$ pairs of columns \emph{not} encoding end-points (for a given row) will always cost
1 flip each to fix. Then either 2 or 0 of the 4 ``endpoint-encoding'' entries will also need to
be flipped; 4 flips will never be necessary because then the row could move to the other haplotype,
requiring no flips. Ungapped-MEC thus maximises the number of rows which require $|V|-2$ rather than
$|V|$ flips. If we think of
$H_1$ and $H_2$ as encoding a partition of the vertices of $V$ (i.e. a vertex i is on one side of the
partition if $H_1$ has 1s in columns $2i-1$ and $2i$, and on the other side if $H_2$ has 1s in those columns), it follows that each row requiring $|V|-2$ flips
corresponds to a cut-edge in the vertex partition defined by $H_1$ and $H_2$. Equation \ref{maxcut} follows.
\begin{flushright}
$\Box$
\end{flushright}
\end{proof}
\textbf{Comment - a rediscovered open problem?}\\
\\
Consider the closely-related ``witness'' version of the (general) MEC problem:\\
\\
\textbf{Problem:} \emph{Witness-MEC}\\
\textbf{Input:} An SNP matrix $M$.\\
\textbf{Output:} For an input matrix $M$ of size $n \times m$, two haplotypes $H_1, H_2 \in \{0,1\}^m$ minimising:
\begin{equation}
D(H_1, H_2) = \sum_{\text{rows }r \in M} \min( d(r,H_1), d(r, H_2) )
\end{equation}
Owing to space restraints we do not prove this here but Witness-MEC is polynomial-time
interreducible with the non-witness, ``counting'' variant (that we have just shown is NP-hard.)\footnote{This result will appear in a forthcoming technical report.}  We mention this because,
when expressed as a witness problem, it can be seen that MEC is in fact a specific type of \emph{clustering} problem. Namely, we
are trying to find two representative ``median'' (or ``consensus'') strings such that the sum, over all input strings,
of the distance between each input string and its nearest median, is minimised. Clustering problems come in many different flavours
and the same problem often reappears, under different guises, in multiple different branches of computer
science. (For example:- information/communication theory, artificial intelligence, computational geometry, string processing,
and data mining.) Related to this, let us define a further problem:\\
\\
\textbf{Problem:} \emph{Binary-Witness-MEC}\\
\textbf{Input:} An SNP matrix $M$ that does not contain any holes\\
\textbf{Output:} As for Witness-MEC\\
\\
What is the complexity of this problem?\footnote{The witnessing and counting versions of this problem
are also polynomial-time interreducible.} Various papers claim that this problem is NP-hard. As far
as we can tell all such claims ultimately lead back to the seminal paper \emph{Segmentation Problems} by
Kleinberg, Papadimitriou, and Raghavan (KCP) \cite{kleinberg}. This paper appears to treat Binary-Witness-MEC
under the guise of the 2-cluster, hypercube variant of
KCP's Segmentation Problem. However, there are two caveats. Firstly, no NP-hardness reduction is given for this
case. Secondly, and more fundamentally, the KCP variant of the problem does not
restrain the rows of the input matrix $M$ like Binary-Witness-MEC does. Specifically, KCP only restricts
the ``decision vectors'' (i.e. the output haplotypes) to the alphabet $\{0,1\}$, while allowing arbitrary
``cost vectors'' (i.e. the rows of the input matrix) from $\mathbb{R}$, a level of freedom
that our problem does not permit.\footnote{Curiously, some variants of KCP's paper \emph{do} discuss a version
where the input matrix is restrained to being binary \cite{kleinbergEco}, but again without proof, and based on the same
foundations as \cite{kleinberg}.} This extra degree of freedom - particularly the ability to simultaneously
use positive, negative and zero values in the input matrix - is what provides the ability to encode
NP-hard problems.\\
\\
If these observations are correct then the complexity of Binary-Witness-MEC and its non-witness counterpart
remain open. From an approximation viewpoint the problem has been quite well-studied;
the problem has a \emph{Polynomial Time Approximation Scheme} (PTAS) because it is a special form of the
\emph{Hamming 2-Median Clustering Problem}, for which a PTAS is demonstrated in \cite{li}. Other approximation results appear
in \cite{kleinberg}, \cite{alon}, \cite{kleinberg2004}, \cite{geometric} and a heuristic for a similar (but
not identical) problem appears in \cite{fasthare}.\\
\\
Finally, it may also be relevant that - as far as we know - the ``geometric'' version of the problem (which uses Euclidean distance rather
than Hamming distance) is also open from a complexity viewpoint. (The version using Euclidean-distance-squared
is NP-hard \cite{drineas}.)
\section{Longest Haplotype Reconstruction (LHR)}
\label{sec:lhr}
\setcounter{equation}{0}
Suppose an SNP matrix $M$ is feasible. Then we can partition the rows of $M$ into two sets, $M_l$ and $M_r$,
such that the rows within each set are pairwise non-conflicting. (The partition might not be unique.)
From $M_i$ ($i \in \{l,r\}$) we can then build a haplotype $H_i$
by combining the rows of $M_i$ as follows: The $j$th column of $H_i$ is set to 1 if at least one row from $M_i$ has a 1 in column $j$,
is set to 0 if at least one row from $M_i$ has a 0 in column $j$, and is set to a hole if all
rows in $M_i$ have a hole in column $j$. Note that, in contrast to MEC, this can lead to haplotypes that
potentially contain holes. For example, suppose one side of the partition contains rows \texttt{10--, -0--} and \texttt{---1};
then the haplotype we get from this is \texttt{10-1}. We define the \emph{length} of a haplotype as the number of positions where it does
not contain a hole; the haplotype \texttt{10-1} thus has length 3, for example. Now, the goal with LHR is to remove
\emph{rows} from $M$ to make it feasible but also such that the sum of the lengths of the two resulting haplotypes is
maximised. We define the function LHR(M) (which gives a natural number as output) as being the
largest value this sum-of-lengths value can take, ranging over all feasibility-inducing row-removals and subsequent partitions.\\
\\
We provide a polynomial-time algorithm for the following variant of LHR:\\
\\
\textbf{Problem:} \emph{Ungapped-LHR}\\
\textbf{Input: } An ungapped SNP matrix $M$\\
\textbf{Output: } The value LHR(M), as defined above.\\
\\
The LHR problem for ungapped matrices was proved to be polynomial time
solvable by Lancia et. al in \cite{lanciabafna}, but only with the genuine restriction that no fragments are included in
other fragments. Our algorithm improves this in the sense that it works for all ungapped input matrices; our
algorithm is similar in style to the algorithm that solves MFR in the ungapped case by Bafna et. al. in \cite{bafna2005}.
The complexity of LHR with gaps is still an open problem. Note that our
dynamic-programming algorithm computes Ungapped-LHR(M) but it can easily be adapted to generate the rows
that must be removed (and subsequently, the partition that must be made) to achieve this maximum.

\smallskip

\begin{lemma}
Ungapped-LHR can be solved in time $O(n^{2}m + n^{3})$
\end{lemma}
\begin{proof}
Let $M$ be the input to Ungapped-LHR, and assume the matrix has
size $n \times m$. For row $i$ define $l(i)~$as the
leftmost column that is not a hole and define $r\left( i\right) $ as the
rightmost column that is not a hole. The rows of $M$ are ordered such that $%
l(i)\leq l(j)$ if $i<j$. Define the matrix $M_{i}$ as the matrix consisting
of the first $i$ rows of $M$ and two extra rows at the top: row $0$ and row $%
-1$, both consisting of all holes. Define $OK(i)$ as the set of rows $j<i$
that are not in conflict with row $i$.

\smallskip

For $h,k\leq i$ and $h,k\geq -1$ and $r(h)\leq r(k)$ define $D[h,k;i]$ as
the maximum sum of lengths of two haplotypes such that:-

\begin{itemize}
\item each haplotype is a combination of rows from $M_{i}$

\item each row from $M_{i}$ can be used to build at most one haplotype (i.e. it cannot be used
for both haplotypes)

\item row $k$ is one of the rows used to build a haplotype and among such rows maximizes $%
r(\cdot )$

\item row $h$ is one of the rows used to build the other haplotype (than $k$) and among such rows
maximizes $r(\cdot )$
\end{itemize}

\smallskip

The solution of the problem $LHR(M)$ is given by

\smallskip 

\begin{equation}
\max_{h,k|r(h)\leq r(k)}D[h,k;n]
\end{equation}

\smallskip

We distinguish three different cases in the calculation of
the $D[h,k;i]$. The first case is when $h,k<i$. Under these
circumstances,

\smallskip 

\begin{equation}
D[h,k;i]=D[h,k;i-1]
\end{equation}

\smallskip 

This is because:-

\begin{itemize}
\item If $r(i)>r(k)$: $i$ cannot be used for the same haplotype as $k$
because $k$ has maximal $r(\cdot )$ among all rows that are used for a
haplotype

\item If $r(i)\leq r(k)$: $i$ cannot increase the length of this haplotype
(because also $l(i)\geq l(k)$)

\item the same arguments hold for $h$
\end{itemize}

\smallskip

The second case is when $h=i$. In this case:

\smallskip 

\begin{equation}
D[i,k;i]=\max_{\substack{ j\in OK(i),\text{ \ ~}j\neq k \\ r(j)\leq r(i)}}%
D[j,k;i-1]+r(i)-\max \{r(j),l(i)-1\}
\end{equation}

\smallskip 

This results from the following. The definition of $D[i,k;i]$ says that row $%
i$ has to be used for the other haplotype than $k$ and amongst such rows
maximizes $r(\cdot )$. Therefore the maximum sum of lengths is achieved by
adding row $i$ to the optimal solution with the restriction that row $j$ is
the most-right-ending row, for some $j$ that agrees with $i$, is not equal
to $k$ and ends before $i$. The term $r(i)-\max \{r(j),l(i)-1\}$ is the
increase in length of the haplotype if row $i$ is added.

\smallskip

The last case is when $k=i$:

\smallskip 

\begin{equation}
D[h,i;i]=\max_{\substack{ j\in OK(i),\text{ \ ~}j\neq h  \\ r(j)\leq r(i)}}%
\left\{ 
\begin{array}{cc}
D[j,h;i-1]+r(i)-\max \{r(j),l(i)-1\} & \text{if }r(h)\geq r(j) \\ 
D[h,j;i-1]+r(i)-\max \{r(j),l(i)-1\} & \text{if }r(h)<r(j)%
\end{array}%
\right.
\end{equation}

\smallskip $\smallskip $

The time for calculating all the $OK(i)$ is $O(n^{2}m)$. When all the $OK(i)$
are known, it takes $O(n^{3})$ time to calculate all the $D[h,k;i]$. This is
because we need to calculate $O(n^{3})$ values $D[h,k;i]$ $(h,k<i)$ that
take $O(1)$ time each and $O(n^{2})$ values $D[i,k;i]$ and also $O(n^{2})$
values $D[h,i;i]$ that take $O(n)$ time each. This leads to an overall time
complexity of $O(n^{2}m+n^{3})$.
\begin{flushright}
$\Box$
\end{flushright}
\end{proof}

\section{The Pure Parsimony Haplotyping Problem (PPH)}
\setcounter{equation}{0}
We refer the reader to Section \ref{intro:pphp} for definitions.\\
\\
\textbf{Problem: } \emph{2-ambiguous Pure Parsimony Haplotyping Problem}\\
\textbf{Input: } A set $G$ of genotypes such that no genotype has more than 2 ambiguous positions\\
\textbf{Output: } $PPH(G)$, which is the smallest number of haplotypes that can be used to resolve $G$.
\begin{lemma}
The 2-ambiguous Pure Parsimony Haplotyping Problem can be solved in polynomial-time.
\end{lemma}
\begin{proof}
We let $n=|G|$ denote the number of genotypes in $G$ and let $m$ denote the length of each genotype
in $G$. We will compute the solution, $PPH(G)$, by
reduction to the polynomial-time solvable problem $MaxBIS$,
which is the problem of computing the cardinality of the
maximum independent set in a bipartite graph.\\
\\
First, some notation. 
A genotype is $i$\emph{-ambiguous} if it contains $i$ ambiguous positions. Each genotype
in $G$ is thus either 0-ambiguous, 1-ambiguous, or 2-ambiguous. For a 0-ambiguous genotype g, we define
$h_{g}$ as the string $g$. For a 1-ambiguous genotype $g$ we let $h_{g:0}$ (respectively, $h_{g:1}$) be the haplotype obtained
by replacing the ambiguous position in $g$ with 0 (respectively, 1). For a 2-ambiguous genotype $g$ we
let $h_{g:i,j}$ - where $i,j \in \{0,1\}$ - be the haplotype obtained by replacing the first (i.e. leftmost) ambiguous position
in $g$ with $i$, and the second ambiguous position with $j$. A haplotype is said to have
even (odd) parity iff it contains an even (odd) number of 1s.\\
\\
Now, observe that there are two ways to resolve a 2-ambiguous genotype $g$: (1) with haplotypes $h_{g:0,0}$ and
$h_{g:1,1}$ and (2) with $h_{g:0,1}$ and $h_{g:1,0}$. Note that - depending on $h$ - one of the
ways uses two \emph{even} parity haplotypes, and the other uses two \emph{odd} parity haplotypes.\\
\\
We build a set $H$ of haplotypes by stepping through the list
of genotypes and, for each genotype, adding the 1, 2 or 4 corresponding haplotypes to the set $H$. (Note
that, because $H$ is a set, we discard duplicate haplotypes.) That is, for
a 0-ambiguous genotype $g$ add $h_g$, for a 1-ambiguous genotype $g$ add $h_{g:0}$ and $h_{g:1}$, and for a
2-ambiguous genotype $g$ add $h_{g:0,0}, h_{g:0,1}, h_{g:1,0}$ and $h_{g:1,1}$.\\
\\
We are now ready to build a bipartite graph $B = (V, E)$ as follows, where $V$ has bipartition
$V^{+} \cup V^{-}$. For each $h \in H$ we introduce a vertex, which we also refer to as $h$; all $h$ with
even parity are put into $V^{+}$ and all $h$ with odd parity are put into $V^{-}$. For each 0-ambiguous genotype $g \in G$ we introduce
a set $I_0(g)$ of four vertices and we connect each vertex in $I_0(g)$ to $h_{g}$. For each 1-ambiguous
genotype $g \in G$ we introduce two sets of vertices $I_1(g,0)$ and $I_1(g,1)$, both containing two
vertices. Each vertex in $I_1(g,0)$ is connected to $h_{g:0}$ and each vertex in $I_1(g,1)$ is connected
to $h_{g:1}$. Finally, for each 2-ambiguous $g \in G$ we introduce (to $V^{+}$ and $V^{-}$ respectively) two sets of vertices $I_2(g,+)$ and
$I_2(g,-)$, each containing 4 vertices. We connect every vertex in $I_2(g,+)$ to every vertex in $I_2(g,-)$,
connect every vertex in $I_2(g,+)$ to the two odd parity haplotypes resolving $g$, and connect
every vertex in $I_2(g,-)$ to the two even parity haplotypes resolving $g$. This completes the construction of $B$.\\
\\
A maximum-size independent set (MIS) of $B$ is a largest set of mutually non-adjacent vertices of $B$. Observe
that, in a MIS of $B$, all the vertices of $I_0(g)$ must be in the MIS, for all 0-ambiguous $g$. To see this,
suppose there exists a 0-ambiguous $g$ such that at least one of the vertices in $I_0(g)$ is not in the MIS.
This is not possible. Firstly, note that if at least one vertex of $I_0(g)$ is in the $MIS$, we should put all of
$I_0(g)$ in the MIS. Secondly, suppose all the vertices in $I_0(g)$ are out of the MIS, but $h_{g}$ is in the
MIS. Then we could simply remove $h_{g}$ from the MIS and add in all the vertices of $I_0(g)$, leading to
a larger MIS:- contradiction! By a similar argument we see that, for all 1-ambiguous $g \in G$, all of
$I_1(g,0)$ and $I_1(g,1)$ must be in the MIS. Now, consider $I_2(g,+)$ and $I_2(g,-)$, for all 2-ambiguous
$g \in G$. We argue that either $I_2(g,+)$ is wholly in the MIS, or $I_2(g,-)$ is wholly in the MIS.
Suppose, by way of argument, that there exists a $g$ such that both $I_2(g,+)$ and $I_2(g,-)$ are
completely out of the MIS. If we are (wlog) free to add all the vertices in $I_2(g,+)$ to the MIS
we have an immediate contradiction. So $I_2(g,+)$ is prevented from being in the MIS by the
fact that one or two of the haplotypes to
which it is connected are already in the MIS. But we could then build a bigger MIS by removing
those (at most) two haplotypes from the MIS and adding the four vertices $I_2(g,+)$; contradiction!\\
\\
We can think of the presence of an $I$ set in the MIS as denoting
that the genotype it represents is resolved using the haplotypes to which it is attached.
Hence, every haplotype that is used for at least one resolution will \emph{not}
be in the MIS, and unused haplotypes \emph{will} be in the MIS. Hence, a MIS will try and minimise the
number of haplotypes used to resolve the given genotypes. Thus:-
\begin{equation}
MaxBIS(B) = 4n + (|H| - PPH(G))
\end{equation}
We can thus use a polynomial-time algorithm for MaxBIS to compute PPH(G).
\begin{flushright}
$\Box$
\end{flushright}
\end{proof}
\textbf{Running time}\\
\\
The above algorithm can be implemented in time $O(mn\log(n)+n^{3/2})$.\\
\\
First we build the graph $B$. We can without too much trouble build a graph representation of $B$ - that combines adjacency-matrix and adjacency-list
features - in $O( mn\log(n) )$ time. For each $g \in G$, add its corresponding $I$ set(s) and add the (at most)
4 haplotypes corresponding to $g$, without eliminating duplicates, and at all times efficiently maintaining adjacency information. Then
sort the list of haplotypes and eliminate duplicate haplotypes (by merging their adjacency information into one
single haplotype.) It is not too difficult to do this in such a way that, in the final data structure representing the graph,
adjacency queries can be answered, and
adjacency-lists returned, in $O(1)$ time. This whole graph construction process takes $O(mn\log(n))$ time.\\
\\
A maximum independent set in a bipartite graph can be constructed from a maximum matching. A maximum matching
in $B$ can be found in time $O(n^{3/2})$ because, in our case, $|V|=O(n)$ and $|E|=O(n)$ \cite{hopcroftkarp}. Once the maximum matching is
found, it needs $O(|E|+|V|)$ time to find a maximum independent set \cite{gavril}. Thus finding a maximum
independent set takes $O(n^{3/2})$ time overall.

\end{document}